\documentclass[aps,prd,reprint,preprintnumbers,floats,epsfig,nofootinbib,amssymb,twocolumn,====
nofootinbib]{revtex4}

\usepackage{slashed}
\usepackage{comment}
\usepackage{graphicx,color}
\usepackage{epsfig}
\usepackage{subfigure}
\usepackage{epsfig}
\usepackage{epstopdf}
\usepackage{dcolumn}
\usepackage{bm}
\usepackage{color}
\usepackage{ulem}
\usepackage{enumitem}
\preprint{ACFI-T21-04}
\usepackage[colorlinks,
            linkcolor=black,
            anchorcolor=black,
            citecolor=black
            ]{hyperref}

\newcommand{\mrm}[1]{\textcolor{magenta}{#1}}

\begin{document}

\baselineskip=15pt

%\preprint{***}

\title{ CP violating dark photon kinetic mixing and Type-III Seesaw}

\author{Yu Cheng$^{1,2}$\footnote{chengyu@sjtu.edu.cn},
Xiao-Gang He$^{1,2,3}$\footnote{hexg@phys.ntu.edu.tw},
Michael J. Ramsey-Musolf$^{1,2,4,5}$\footnote{ mjrm@sjtu.edu.cn, mjrm@physics.umass.edu},
Jin Sun$^{1,2}$\footnote{019072910096@sjtu.edu.cn}}

\affiliation{${}^{1}$Tsung-Dao Lee Institute \& School of Physics and Astronomy, Shanghai Jiao Tong University, Shanghai 200240, China}
\affiliation{${}^{2}$Shanghai Key Laboratory for Particle Physics and Cosmology, Key Laboratory for Particle Astrophysics and Cosmology (MOE), Shanghai Jiao Tong University, Shanghai 200240, China}
\affiliation{${}^{3}$Department of Physics, National Taiwan University, Taipei 10617, Taiwan}
\affiliation{${}^{4}$Amherst Center for Fundamental Interactions, Department of Physics, University of Massachusetts, Amherst, MA 01003, USA, USA}
\affiliation{${}^{5}$Kellogg Radiation Laboratory, California Institute of Technology, Pasadena, CA91125, USA}

\begin{abstract}
The hypothetical dark photon portal connecting the visible and dark sectors of the Universe has received considerable attention in recent years, with a focus on CP-conserving kinetic mixing between the Standard Model (SM) hypercharge gauge boson and a new U(1)$_X$ gauge boson. In the effective field theory context, one may write down non-renormalizable CP-violating kinetic mixing interactions involving the $X$ and SU(2)$_L$ gauge bosons.
We construct for the first time a renormalizable model for CP-violating kinetic mixing that induces CP-violating non-Abelian kinetic mixing at mass dimension five.
The model grows out of the type-III seesaw model, with the lepton triplets containing right-handed neutrinos playing a crucial role in making the model renormalizable and providing
a bridge to the origin of neutrino mass. This scenario also accommodates
electron  electric dipole moments (EDM) as large as current experimental bound, making future EDM searches an important probe of this scenario.

\begin{comment}
{\color{blue}
Dark photon physics has attracted significant theoretical and experimental attention.
A dark photon from a $U(1)_X$ gauge group can have renormalizable kinetic
mixing with the $U(1)_Y$ gauge boson of the SM $SU(3)_C\times  SU(2)_L\times U(1)_Y$ gauge group. It only permits CP conserving mixing. When a $SU(2)_L$ triplet scalar with a non-zero vacuum expectation value is introduced, a dimension 5 operator can be constructed to generate a CP violating
kinetic mixing. To further realize a renormalizable model for CP violating kinetic mixing, new ingredients are needed. We construct for the first time such a model.
We find that the triplet leptons in type-III seesaw model can play the crucial role in building such a renormalizable model. This natural connection of dark photon and neutrino physics may have important implications about how new physics might emerge. The CP violating kinetic mixing induced electron electric dipole moment (EDM) can be as large as current experimental bounds. The model can be further tested by improved measurement of electron EDM.
}
\end{comment}
\end{abstract}

\maketitle
The dark sector in our universe, for which dark matter and dark energy provide the primary evidence, remains largely unexplained.
The dark sector may be well-described by simple field content with primarily gravitational interactions. However, there exists a plethora of theoretical proposals for a richer dark sector containing multiple particles and new interactions. These possibilities include interactions between the dark sector and Standard Model (SM)  of particle physics, referred to as portals. The most widely considered include the Higgs, axion, neutrino, and dark photon portals, each of which implies distinctive phenomenological consequences.
Here, we focus on the novel possibility of CP-violating (CPV) dark photon portal interactions.

\begin{comment}
{\color{blue}
Searching for new physics beyond the Standard Model (SM) \mrm{of particle physics}is always the frontier of fundamental physics research. Because of null search results for new physics beyond the SM (BSM) at the Large Hadron Collider (LHC), other possibilities have received increased attention in recent years. A hypothetical dark photon and related physics phenomena at both the energy and intensity frontiers are among the very interesting possibilities. A dark photon linking a possible dark sector through kinetic
mixing can act as a portal to reveal many interesting BSM phenomena  and related studies have attracted considerable
attention in theoretical model building and experimental searches, as well as in astrophysics and cosmology investigations.  To better understand
the implications of dark photon, an in-depth exploration of the dark photon itself and its kinetic mixing portal  are of paramount importance.
}
\end{comment}

To date, most portal studies - dark photon or otherwise - have focused on CP-conserving interactions.
While of interest in its own right, the new CPV interactions beyond those of the SM are also needed to explain the cosmic matter-antimatter asymmetry.
In the case of the dark photon portal, it is straightforward to construct a CP-conserving portal. Indeed, it has long been realized that a dark photon gauge field  $X_\mu$ associated with a beyond SM $U(1)_X$ gauge group can mix with the $U(1)_Y$ gauge field $B$ in the SM gauge group $SU(3)_C\times SU(2)_L\times U(1)_Y$ through a renormalizable kinetic mixing term\cite{Galison:1983pa,Holdom:1985ag,Foot:1991kb}: $X_{\mu\nu} B^{\mu\nu}$.
One may write down a  CPV
$\tilde X_{\mu\nu}B^{\alpha\beta}$ term.  Here $X_{\mu\nu} = \partial_\mu X_\nu - \partial_\nu X_\mu$ and $\tilde X_{\mu\nu} = \epsilon_{\mu\nu\alpha\beta} X^{\alpha\beta}/2$. However, this interaction has no physical effect at the perturbative level because it can be written as a total derivative proportional to $ \partial_{\mu}(\epsilon^{\mu\nu\alpha\beta}X_\nu \partial_\alpha B_\beta)$.

In extended models, a dark photon may also mix kinetically
with the  SM non-abelian $SU(2)_L$  gauge bosons~\cite{Chen:2009ab,Chen:2009dm,Barello:2015bhq,Arguelles:2016ney,Fuyuto:2019vfe}.
\begin{comment}
{\color{blue}
 A $U(1)_X$ gauge field $X$ which may not have direct interactions with SM particle is a typical dark photon. However, it has long been realized that $X$ can mix with the $U(1)_Y$ gauge field $B$ in the SM gauge group $SU(3)_C\times SU(2)_L\times U(1)_Y$ through a renormalizable kinetic mixing term $X_{\mu\nu} B^{\mu\nu}$. This kinetic mixing acts as a portal communication between
 the dark sector and our visible world~\cite{Galison:1983pa,Holdom:1985ag,Foot:1991kb}. In extended models, a dark photon can also have kinetic mixing with the non-abelian $SU(2)_L$ in the SM~\cite{Chen:2009ab,Chen:2009dm,Barello:2015bhq,Arguelles:2016ney,Fuyuto:2019vfe}. The usual kinetic mixing term is a CP conserving interaction. One may write down a nominally CPV
$\epsilon^{\mu\nu \alpha\beta} X_{\mu\nu}B_{\alpha\beta}$ term.  However this interaction has no physical effect at the perturbative level because it can be written as a total derivative proportional to $\partial_{\mu}(\epsilon^{\mu\nu\alpha\beta}X_\nu \partial_\alpha B_\beta)$.
}
\end{comment}
In this context, it has been shown recently that CPV kinetic
mixing between a dark photon and SM gauge particles can arise~\cite{Fuyuto:2019vfe}.
If one includes a zero hypercharge $SU(2)_L$ Higgs triplet, one may construct a
(non-renormalizable) dimension five operator containing such a term.
Several interesting phenomenological consequences follow, notably a new CPV source for electric dipole moments (EDMs) of SM fermions, and possible collider signature in jets angular distributions.

However, a renormalizable model realization of this possibility has thus far been lacking.
The presence of a non-renormalizable interaction implies the existence of new particles and interactions whose detailed nature is not evident from the structure of the low-energy effective operator alone. For example, existence of the well-known dimension five neutrino Majorana mass operator would imply non-conservation of total lepton number at the classical level without revealing its fundamental origin.  The construction and phenomenology of models ({\it e.g.} the see saw mechanism) generating this interaction have received intensive theoretical interest over the years.
In a similar spirit, we construct here for the first time a renormalizable model with CPV kinetic mixing between a dark photon  and  $SU(2)_L$ gauge bosons and
analyze the implications for future EDM searches. We also give general considerations for building such a model, which may be realized in other constructions.

Two key minimal ingredients are needed for this purpose: (i) The $SU(2)_L$ Higgs triplet~\cite{Fuyuto:2019vfe}  $\Sigma^a$: $(1,3)(0, 0)$,  where the brackets denote the transformation properties under $SU(3)_C\times SU(2)_L$ and $U(1)_Y\times U(1)_X$ gauge symmetries, respectively. The neutral component $\Sigma^0$ obtains a non-zero vacuum expectation value (vev) $<\Sigma^0> = v_\Sigma$\footnote{Replacing $\Sigma^a$ by a composite triplet, such as  $H^\dagger \tau^a H$ from the Higgs doublet H~\cite{Barello:2015bhq},  a dimension six kinetic mixing operator can also be generated.}. $\Sigma^a$ is needed so that the $SU(2)_L$ index ``a'' of the gauge triplet field $W^a$ can be contracted to
form a gauge group singlet dimension five operator
\begin{equation}
\label{eq:nonren}
\epsilon^{\alpha\beta \mu\nu}X_{\alpha\beta}W^a_{\mu\nu} \Sigma^a \ \ \ ,
\end{equation}
which is non-renormalizable.  (ii) introduction of new fields $f$ that, when integrated out, yield the interaction Eq.(\ref{eq:nonren}).
$f$ transforms as : $(1, n)(0, x_f)$ and cannot be an $SU(2)_L$ singlet $n=1$ in order to mix $W$ and $X$.  The source of CPV depends on how $\Sigma$ interacts with $f$.
{\it A priori}, the fields may be fermions or scalars\footnote{One may also want to consider a vector boson running in the loop. Since we are working with renormalizable theory, vector particle if not gauge particle, may complicate the model building. We will not venture into this possibility.}. However, $f$ cannot be a scalar since the tensor $\epsilon^{\alpha\beta \mu\nu}$ cannot arise from tree-level scalar exchange built from renormalizable interactions nor from scalar loops. It can, however,  arise from loops containing a chiral fermion $f$ with $\gamma_5$ and a CPV interaction appearing in the  $f{\bar f} \Sigma$ couplings (see further discussion below).
Since $f$ is a chiral fermion, one needs to pay attention to make sure the model is gauge anomaly free.
The minimal $SU(2)_L$ representation would be $n=2$.
Stringent limits from fractionally charged particle searches\cite{Zyla:2020zbs}
imply that the components of $f$ must have integer charges. To have integer electric charges for the components in $f$, it requires that the hypercharge for $f$ must be a half integer.
In this case, exchange of $f$ in the loop will not only generate $X-W$, but also $Y-X$ and $Y-W$ kinetic mixing terms.
The simplest choice is actually $n=3$ with zero hypercharge.
The components in $f$ have zero or $\pm1$ electric charge.
 Taking the triplet to be right-handed, $f = f_R$, makes it possible to facilitate the type-III seesaw model~\cite{type3-seesaw} for neutrino mass generation by identifying the neutral component in  $f_R$ as the heavy right-handed  neutrino. 

If one makes the theory supersymmetric, the LSP of the model may also provide a dark matter candidate.
Here we will concentrate on the non-supersymmetric model for purposes of simplicity and illustration. Other choices for the mediator particle content -- satisfying the aforementioned criteria -- may have distinctive phenomenological consequences and connections with other open problems in particle physics and cosmology.
We will see later that our triplet mediator model will produce a fermion electric dipole moment as unique signature for CPV kinetic mixing.

We will assign one of $f_R$, denoted $f_1$, to transform as $(1, 3)(0, x_f)$.
Since $f_1$ is a chiral field, one must include more than one such multiplet with different $x_f$ charges in order to ensure anomaly cancellation.
To this end, we  introduce a second  $f_R$, $f_2: (1,3)(0, - x_f)$, whose contribution to the anomaly will cancel that from $f_1$.
If future experiments indicate non-zero masses for all three light neutrinos (one massless neutrino is consistent with present neutrino oscillation data), inclusion of a third $f_R$ would be necessary: $f_3: (1,3)(0,0)$ which
does not generate any gauge anomaly. $f_3$ is inessential for our purposes.
The component in  $f_R$ can be written as
\begin{eqnarray}
&&f_R = {1\over 2}\sigma^a f^a_R = {1\over 2} \left (\begin{array}{cc}
f^0_R&\;\;\sqrt{2}f^+_R\\
\\
\sqrt{2}f^-_R&\;\;-f_R^0
\end{array}
\right )\;,
\end{eqnarray}
and $f_L = f^c_R\;  (f^+_L = (f^-_R)^c, f^0_L=(f^0_R)^c, f^-_L = (f^+_R)^c)$.

After integrating out the $f$ fields, the dimension five operator required is given by
\begin{equation}
\label{eq:CPVop}
\mathcal{L}_X = - (\tilde \beta_X / \Lambda) \mathrm{Tr}(W_{\mu\nu}\Sigma)\tilde
X^{\mu\nu}.
\end{equation}
Expanding it, we have
\begin{eqnarray}
\mathcal{L}_X&\to&-{\tilde \beta_X \over 2 \Lambda} \tilde X^{\mu\nu} \left [ (s_W F_{\mu\nu} +c_W Z_{\mu\nu}) \right .\nonumber\\
&&\left . + i g(W^-_\mu W^+_\nu - W^+_\mu W^-_\nu) \right ](v_\Sigma + \Sigma^0)\;. \label{cpv}
\end{eqnarray}
The same loop integral also generates the CP conserving counterpart
$ - (\beta_X / \Lambda) \mathrm{Tr}(W_{\mu\nu}\Sigma)X^{\mu\nu}$ whose expanded form is obtained by replacing $\tilde X^{\mu\nu}$ by $X^{\mu\nu}$ in the above.
Here we have normalized the fields as
\begin{eqnarray}
W_\mu =
{1\over 2} \left ( \begin{array}{cc}
W^0_\mu&\sqrt{2}W^+_\mu\\
\sqrt{2}W^-_\mu&-W^0_\mu
\end{array}
\right ),\Sigma
={1\over 2} \left ( \begin{array}{cc}
\Sigma^0&\sqrt{2}\Sigma^+\\
\sqrt{2}\Sigma^-&-\Sigma^0
\end{array}
\right ),
\end{eqnarray}
where $W^0_\mu$ is a linear combination of the photon $A_\mu$ and Z field
$Z_\mu$ with $ W^0_\mu = \sin\theta_W A_\mu + \cos\theta_W Z_\mu$.

To make the dark photon mass $m_X$ non-zero, we introduce a scalar $S_X: (1,1)(0, -2x_f)$  with a vev $<S_X>=v_s/\sqrt{2}$. We obtain $m^2_X = x^2_f g^2_X v^2_s$. It also contributes to heavy neutrino masses.

\begin{figure}
	\setlength{\abovecaptionskip}{-20pt}
	\centering
	\includegraphics[width=.35\textwidth]{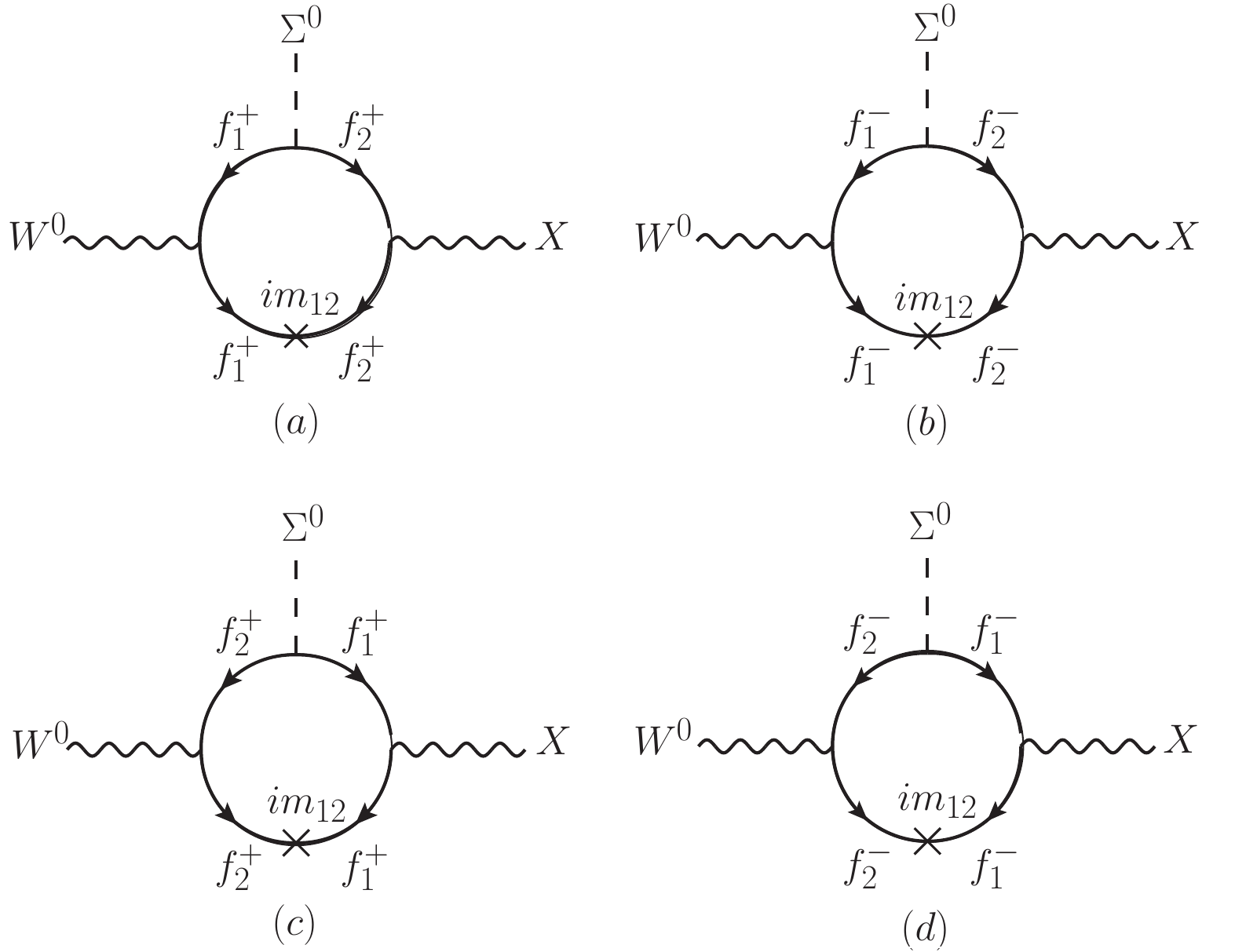}
	\vspace{1cm}
	\caption{The one loop diagrams contributing to kinetic mixing.}
	\label{fig1}
\end{figure}

We now discuss how to generate a non-zero  $\tilde \beta_X$.
The one loop Feynman diagrams are shown in Fig.\ref{fig1}.
The coupling of $\Sigma$ to $f$  is crucial to the model and is given by
\begin{eqnarray}
 4 \mathrm{Tr}(\bar f^c_{R\;i}Y_{f \sigma} \Sigma f_{R\;j}) =  iY_{f \sigma } \bar f^a_{L\;i} \Sigma^b f^c_{R\;j} \epsilon^{abc}\,.
\end{eqnarray}
The appearance of $\epsilon^{abc}$ requires
 more than one $f$.
The couplings between $\Sigma_0$ and $f$ needed in Fig.~\ref{fig1}, are given by
\begin{eqnarray}
&& Y_{f\sigma12} \left ( \overline{( f^+_{R\;1})^c }f^+_{R\;2} - \overline{( f^-_{R\;1})^c} f^-_{R\;2}\right ) (v_\Sigma + \Sigma^0)\;.
\end{eqnarray}

The other Yukawa couplings terms for leptons and quarks responsible to masses are given by
\begin{eqnarray}
&&- \bar L_L Y_e \tilde H E_R - \bar L_L Y_{fL3} \tilde H f_{R\;3}
- \bar f^c_{R\;1} Y_{fs1} S_X f_{R\;1} \nonumber\\
&&- \bar f^c_{R\;2} Y_{fs2} S^\dagger_X f_{R\;2} - \bar f^c_{R\;1} m_{12}
f_{R\;2} - \bar f^c_{R\;3} m_{33} f_{R\;3} \nonumber\\
&& -\bar Q_L Y_u H U_R - \bar Q_L Y_d \tilde H D_R\;.
\end{eqnarray}
Since $f_{1,2}$ do not couple to $H$, the resulting Dirac neutrino mass matrix term $Y_{fL}$ is only rank one
which is not acceptable phenomenologically.
This problem can be cured by introducing additional scalar $SU(2)_L$ doublets
$H'_1: (1,2)(-1/2, -x_f)$ and $H'_2: (1,2)(-1/2, x_f)$, whose vevs are $v'_1/\sqrt{2}$ and $v'_2/\sqrt{2}$, respectively, so that
$- \bar L _LY_{fL 1} H'_1 f_{R\;1} - \bar L _LY_{fL 2} H'_2 f_{R\;2}$ terms can
be added. 
The usual electroweak scale constrains Higgs doublets vevs since $v^2+v'^{2}_1+v'^{2}_2=(246 \mbox{GeV})^2$, due to the bound from W boson mass.  Among the vevs, from naturalness consideration,
the vev providing the top quark mass should be the largest one, therefore $v'_{1,2}$ should be smaller than $v$. However, it does not mean that $v'_{1,2}$ need to be very small. They will contribute to neutrino masses, but the small neutrino masses are not due to smallness of  $v’_{1,2}$ because seesaw mechanism is in effect in our model. 
Lepton mass matrices are given by
\begin{eqnarray}
\mathcal{L}_m =&&-{1\over 2} (\bar \nu_L, \bar \nu^c_R) \left ( \begin{array}{cc}
0&\;\;M_D\\
M_D^T&\;\;M_R
\end{array}
\right ) \left ( \begin{array}{cc}
\nu^c_L\\ \nu_R
\end{array}
\right )\nonumber\\
&& - (\bar E_L, \bar f_L) \left ( \begin{array}{cc}
m_e&\;\;\sqrt{2}M_D\\
0&\;\;M_R
\end{array}
\right ) \left ( \begin{array}{cc}
E_R\\ f_R
\end{array}
\right ),
\end{eqnarray}
where $m_e = Y_e v / \sqrt{2}$ is an arbitrary $3\times 3$ matrix. $M_D$ is a full $3\times 3 $ matrix with the 3 column elements $Y_{fL i 1}v'_1/\sqrt{2}$,  $Y_{fL i 2}v'_2/\sqrt{2}$ and $Y_{fL i 3}v/\sqrt{2}$, respectively. $M_R$ is a symmetric $3\times 3$ matrix with non-zero entries $R_{11} = m_1=Y_{fs1} v_s/ \sqrt{2}$, $R_{22} =m_2= Y_{fs2} v_s/ \sqrt{2}$, $R_{12} = m_{12}$ and $R_{33}= m_{33}$. 
The structure of $M_R$ allows one to separate the seesaw scale represented by $\sim m_{33, 12}$ from the dark $U(1)_X$ breaking scale $\sim v_s$ so that $m_X$ can be smaller than the seesaw scale.
Note that all entries in $M_R$ will break lepton number by two units after $S_X$ develops non-zero vev. The largest element in  $M_R$ will set the seesaw scale.

Expanding the kinetic Lagrangian terms for $f$, $\Sigma$ and $S_X$,
$\mathcal{L}_{K}=2 \mathrm{Tr}(\bar f_{R\;j} i\gamma_\mu D^\mu f_{R\;j}) + 2 \mathrm{Tr}((D_\mu \Sigma)^\dagger (D^\mu \Sigma)) + (D_\mu S_X)^\dagger (D^\mu S_X)$, we have $W^0$ and $X$ couplings with $f_R$ necessary for the  remaining vertices in Fig.~\ref{fig1}, the interaction  Lagrangian $\mathcal{L}_{int}$ is given by
\begin{eqnarray}
&&g_X x_f \left [ Tr(\bar f_{R\;1}\gamma^\mu X_\mu f_{R\;1}) - Tr(\bar f_{R\;2}\gamma^\mu X_\mu f_{R\;2})  \right ]  \\
&&+ g \left [Tr(\bar f_{R\;1}\gamma^\mu W_\mu
f_{R\;1}) + Tr(\bar f_{R\;2}\gamma^\mu W_\mu  f_{R\;2})  \right ] \;.\nonumber
\end{eqnarray}

Evaluating the diagrams in Fig.~\ref{fig1} yields
\begin{eqnarray}
M^{\mu\nu} =&&  {i\over 4\pi^2} g g_X x_f m_{12}Y^{*}_{f\sigma} \epsilon^{\mu\nu\alpha\beta}p_{X\alpha} p_{W\beta} \\&&\times \left (f(m_1,m_2, p_W, p_X) + f(m_2,m_1,p_W, p_X)\right )\;.\nonumber
\end{eqnarray}
where
\begin{eqnarray}
&&f(m_1,m_2, p_W, p_X) =\int^1_0 dx\int^{1-x}_0dy (1-x-y)\\
&&\times \left [ D(m_1,m_2, p_W,p_X)+  D(m_2,m_1,p_X,p_W) \right ]\;,\nonumber
\end{eqnarray}
where $D(m_1,m_2, p_W, p_X) = (m_1^2/(m_1^2-m_2^2))/(m^2_1 - y(m^2_1-m^2_2) -xp^2_W - yp^2_X + ( xp_W-yp_X)^2)$.

Matching $\mathcal{L}_X$ in Eq.~(\ref{cpv}), $2 \epsilon^{\mu\nu\alpha\beta}p_{X\alpha}p_{W\beta} \epsilon_{W\mu} \epsilon^*_{X\nu} \to - \tilde X^{\mu\nu}W^0_{\mu\nu}$,
implies
\begin{eqnarray}
{\tilde \beta_X\over \Lambda} &&= {1\over 2 \pi^2} g g_X x_f \mathrm{Im} (m_{12} Y_{f\sigma}^*)\\
&&\times  \left [ f(m_1,m_2,p_W, p_X) + f(m_2,m_1,p_W, p_X)\right ]\;.\nonumber\label{cpv-mixing}
\end{eqnarray}
The phase $\delta$ of $m_{12} Y_{f\sigma}^* $ is the CPV source.
$\beta_X /\Lambda$ is obtained by replacing $\mathrm{Im}(m_{12}Y_{f\sigma}^*)$ in the above by $\mathrm{Re} (m_{12}Y_{f\sigma}^*)$.
The kinetic mixing in Eq.~(\ref{cpv}) corresponds to $p_W + p_X=0$ and $q^2=(p_W +p_X)^2=0$. Note that the presence of the $\epsilon^{\mu\nu\alpha\beta}$ in (\ref{eq:nonren},\ref{eq:CPVop}) results from taking the trace of the fermion loop with $\gamma_5$ type of coupling for $\Sigma$ to $f$ and the mass mixing term  $m_{12}$. Loops containing scalars or vector-like fermions would not yield this structure.

The CP conserving kinetic mixing terms cause mixing of gauge fields.
We have the following relevant Lagrangian $\mathcal{L}_{KM}$ terms before writing the gauge fields in canonical form,
\begin{eqnarray}
&& -{1\over 4} F_{\mu\nu} F^{\mu\nu} - {1\over 4} Z_{\mu\nu} Z^{\mu\nu} -
{1\over 4} X_{\mu\nu}X^{\mu\nu}
-{1\over 2} \epsilon_{AX} F_{\mu\nu} X^{\mu\nu} \nonumber\\
&&- {1\over 2} \epsilon_{ZX}Z_{\mu\nu} X^{\mu\nu} + {1\over 2} m^2_Z Z_\mu Z^\mu
+ {1\over 2} m^2_X X_\mu X^\mu\;,  \label{LKM}
\end{eqnarray}
where $\epsilon_{AX} = \alpha_{XY} c_W + \beta_X s_W v_\Sigma/\Lambda$ and $\epsilon_{ZX} = -\alpha_{XY}s_W + \beta_X c_W v_\Sigma/\Lambda$ with $\alpha_{XY}$ defined by a possible $U(1)_Y$ and $U(1)_X$ kinetic mixing term $-(1/2)\alpha_{XY} X^{\mu\nu} B_{\mu\nu}$ which is independent of $\beta_X$ . Here we  use the $U(1)_Y$ gauge field $B_\mu = \cos\theta_W A_\mu - \sin\theta_W Z_\mu$.

The mass eigen-states, the photon $A^m$, the Z-boson $Z^m$ and the dark photon $X^m$, to the leading order in small mixing parameters $\epsilon_{AX}$ and $\epsilon_{ZX}$ are related to the original gauge fields $A$, $Z$
and $X$ by
\begin{eqnarray}
\left ( \begin{array}{c}
A\\
Z\\
X
\end{array} \right ) =
\left (\begin{array} {ccc}
1\;\;\;\; &  0\; \;\;\;& - \epsilon_{AX} \\
0\;\;\;\;& 1  \;\;\;\;&  - \xi - \epsilon_{ZX}\\
0\;\;\;\;& \xi\;\;\;\;& 1
\end{array}
\right )
\left ( \begin{array}{c}
A^m\\
Z^m\\
X^m
\end{array}
\right )\;, \label{mixing-matrix}
\end{eqnarray}
where $\xi$ is the angle describing
$X$ and $Z$ mass mixing, $\xi \approx - m^2_Z\epsilon_{ZX}/(m^2_Z - m^2_X)$.
There is an enhancement for $\xi$ when $m_X$ is close to $m_Z$.
The modifications for the interaction terms $J^\mu_{em}A_\mu$, $J^\mu_Z Z_\mu$ and $J^\mu_XX_\mu$ can
be obtained by replacing the fields according to Eq.~(\ref{mixing-matrix}). Later we will drop the superscript $m$ on the mass eigen-state gauge fields.

We now discuss some related consequences. Before doing that let us estimate how large $\tilde \beta_X$ might be from data.  For simplicity, we
consider nearly degenerate case $m_1 \approx m_2 \approx |m_{12}| \approx m >> m_{W,X}$ . In this limit, we have $\tilde \beta_X/\Lambda \approx g g_X x_f  |Y^*_{f\sigma}|\sin\delta /6\pi^2 m  $. The size of $\tilde
\beta_X /\Lambda$ is governed by the seesaw mass scale represented by elements in $M_R$.
The size of $v_\Sigma$ is also important  as can be seen from Eq.~(\ref{cpv}),  which is constrained by $\rho = 1.00038\pm 0.00020$ from electroweak precision tests~\cite{Zyla:2020zbs}. A non-zero $v_\Sigma$ would modify $\rho$ from 1 in SM to $1 + 4v^2_\Sigma/ v^2$.
Therefore data implies $v_\Sigma <3$GeV at the 2$\sigma$ level. Both ATLAS and CMS experiments at the LHC have carried searches for heavy fermion in type-III seesaw model and found the mass to be larger than 790 GeV~\cite{Aad:2020fzq} (880 GeV~\cite{Sirunyan:2019bgz}) at the 95\% C.L..  The seesaw scale should be above this limit.
For purposes of illustration, allowing both  $g_X x_f$ and $Y_{f\sigma}$ to be as large as their perturbative unitarity bound
with approximately  $\sqrt{4 \pi}$, we obtain at 95\% C.L. an upper bound $5 \sin\delta \times 10^{-4}$ for $\tilde \beta_X v_\Sigma /\Lambda$. The bound for $\beta_X$ is obtained by replacing $\sin\delta$ by $\cos\delta$.

In Ref.~\cite{Fuyuto:2019vfe} two interesting CP violating effects due to CP
violating kinetic mixing were identified, the  EDMs of SM fermions and a collider signature in di-jet angular distribution asymmetry. The latter has large background and is difficult to observe.
We find the jet analysis is again difficult to probe in our model, but the effects on electron EDM can be dramatic and provide a good test for the model~\cite{he-neutron,Engel:2013lsa,michael-edm}.

\begin{figure}
	\setlength{\abovecaptionskip}{-20pt}
	\centering
	\includegraphics[width=8cm]{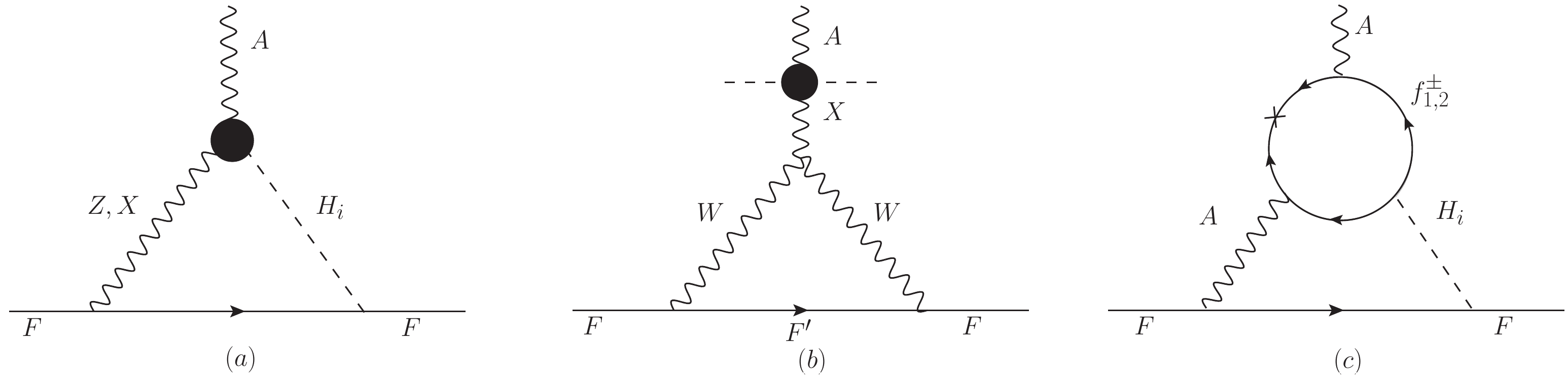}
	\vspace{1cm}
	\caption{Different contributions to fermion EDMs.}
	\label{fig2}
\end{figure}

The EDM $d_F$ for a SM fermion $F$ may be induced by three different contributions shown in Fig.~\ref{fig2}. The contribution from Fig. 2(a) arises in all model constructions, since it contains the contribution from the dimension five operator (\ref{eq:nonren}). It has been studied in Ref.~\cite{Fuyuto:2019vfe}. In the present context, it is actually a two-loop contribution, since $\tilde\beta_X$ is generated at the one-loop level.
Fig.~2(b) first generates an $X$ EDM (replacing external $A$ by $X$). As  seen from Eq.~(\ref{mixing-matrix}) there is no mixing for $X$ to have final mass eigen-state of the photon $A$ for $q^2=0$, so no ordinary EDM for $F$ can be generated. But a weak EDM $d^W_F$ for a fermion through $X$ mixing with $Z$ will be produced.
This however only provides a very weak constraint~\cite{Zyla:2020zbs}.  The third contribution is the Barr-Zee diagram~\cite{barr-zee} shown by Fig. 2 (c). In the diagram, the loop with two photons attached is obtained by replacing $X$ with $W^0$ in Fig.~\ref{fig1}. One finds contributions from Fig.~1(a) and Fig.~1(c) are
cancelled by those from Fig.~1(b) and Fig.~1(d) due to antisymmetric $\Sigma$-f-f coupling in fermion space. This is a generic property of this type of model for higher representations which produce a antisymmetric $\Sigma$-f-f coupling. Had one chosen a different representation, such as $n=2$ for $f$, the contribution from Fig.~2(c) would not vanish.  Its effect will dominate fermion EDM as shown in Ref.\cite{barr-zee} making the test of CPV kinetic mixing signature difficult. This makes our model with $f$ being a triplet unique from experimentally testing of CP violation in kinetic mixing.

Therefore the dominant contribution for $d_F$ is from Fig.~2(a). The expression has been given in Eq.~(11) of Ref.\cite{Fuyuto:2019vfe}. Since there are several Higgs doublets $H_i$, one should replace the mixing
$s_\theta c_\theta f(r_{ZH_1}, r_{ZH_2})$ and $s_\theta c_\theta f(r_{XH_1}, r_{XH_2})$ by  $\Sigma_{i = [1, N-1]} V_{\Sigma i} V_{hi} f(r_{ZH_i}, r_{ZH_\Sigma})$ and  $\Sigma_{i = [1, N-1]} V_{\Sigma i} V_{hi} f(r_{XH_i}, r_{XH_\Sigma})$ with $N=5$. Here the summation sums over all mass eigenstates $H_i$ where the original scalar states $\Sigma^0$ and  the neutral real component $h$ in $H$, are expressed as linear combinations of $H_i$, $\Sigma^0 =\sum_i V_{\Sigma i} H_i$ and $h = \sum_i V_{h i} H_i$.
 The formula in Eq.~(11) of Ref.\cite{Fuyuto:2019vfe} assumed CPV source comes from $\tilde \beta$ which is proportional to the invariant
phase from $m_{12}Y^*_{f\sigma}$. One may also wonder if CPV exists in the vertex $H_i$ to $F$, that is if $V_{\Sigma i} V_{hi}$ which depends on CPV phase in the Higgs potential mix different neutral components from $H_i$, $\Sigma$ and $S_{X}$. We have checked in detail that there is no CPV in the Higgs potential with the quantum numbers assigned for the Higgs bosons.

The electron EDM $d_e$ prediction is constrained directly from experimental bounds, assuming a \lq\lq sole source" analysis of the polar
molecule system~\cite{michael-edm}. For the neutron EDM, we use $d_n =
-0.233 d_u + 0.774 d_d + 0.008d_s$ obtained in Ref.~\cite{Bhattacharya:2015wna}.  For illustration, we assume $V_{\Sigma 1} V_{h 1}$ term dominates the contribution with some benchmark values for the heavy $m_{H_\Sigma}$
and $m_X =60$ GeV and kinetic mixing parameters.
 In the model, $m_{H_\Sigma}$ should be below the largest scale -- the seasaw scale, whose lower limit set by the LHC~\cite{Aad:2020fzq,Sirunyan:2019bgz} is about 900 GeV.
The 13 TeV LHC excludes also a real triplet lighter than 275 (248) GeV for a range of parameter space~\cite{Chiang:2020rcv}. We therefore take benchmark range $300 \sim 600$ GeV for $m_{H_\Sigma}$.
The dependence on $m_X$ is weak except near $m_Z$. Since $m_{H_1}$ is the SM-like Higgs boson, its mass is assumed to be 125 GeV.
For mixing parameters, we consider two cases to show details. {\it Case I:} kinetic mixing parameters are all generated from our loop calculations, that is $\alpha_{XY}=0$. To maximize the contribution to EDMs, we choose $\sin\delta = \cos\delta = 1/\sqrt{2}$.  {\it Case II:} we
choose a large, but allowed $\alpha_{XY}$ to be $10^{-2}$ and set $\tilde
\beta_X/\Lambda$ to the maximal allowed value.

\begin{figure}[!t]
	\centering
        {\includegraphics[width=.45\textwidth]{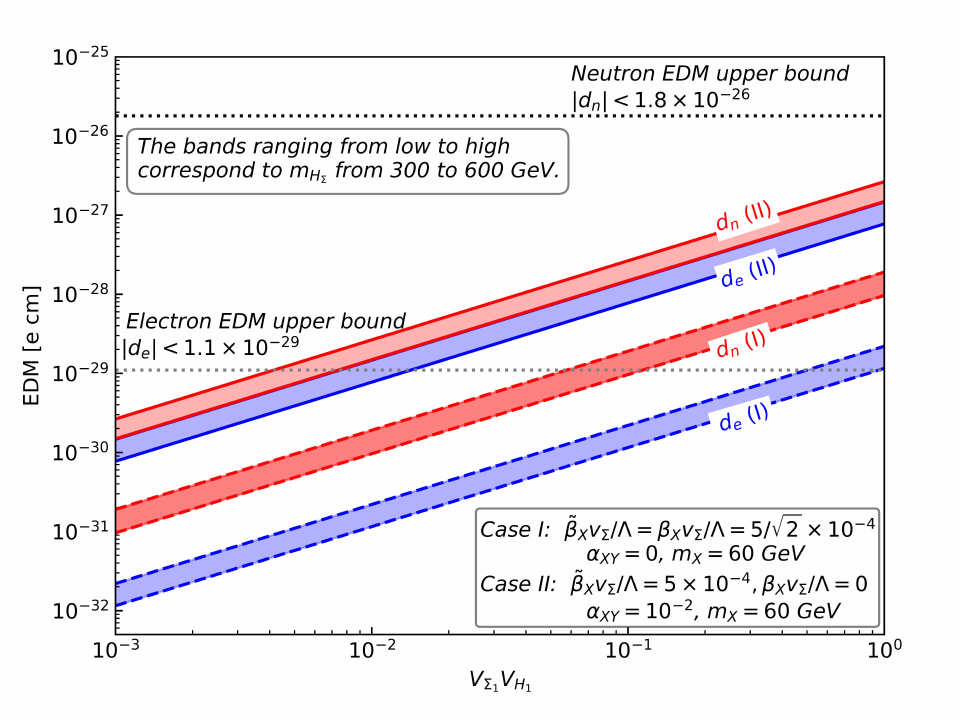}}
	\caption{ The ranges for $d_n$ and $d_e$ for {\it Cases I} and {\it II}.
	The 90\% C.L. upper limits for electron  and neutron EDMs  are from Ref.~\cite{Andreev:2018ayy} and  Ref.~\cite{Abel:2020gbr}, respectively.
}
	\label{EDM2}
\end{figure}

We show the results in Fig.~\ref{EDM2}.
The splitting between $d_e$ and $d_n$ for cases I and II behave differently is because for Case II, the addition of $\alpha_{XY}$ in $\beta$ modifies $d_e$ and $d_n$ differently.
Note that for both {\it Cases I} and {\it II},  EDM is larger with a larger $m_{H_\Sigma}$ because the first term in $f(x,y)$ defined in Ref.\cite{Fuyuto:2019vfe} has a $ln(m^2_{H_\Sigma}/m^2_h)$ term  increasing with $m_{H_\Sigma}$. However, it will not increase indefinitely with $m_{H_\Sigma}$ because the mixing parameter $V_{\Sigma 1} V_{h 1}$ will vanish in that limit.
The current neutron EDM does not constrain the parameters for both cases. But with improved sensitivity, such as nEDM experiment at the Spallation Neutron Source~\cite{Leung:2019wao}
 or n2EDM experiment at the Paul Scherrer Institute~\cite{Ayres:2021hoq}, where a sensitivity of $10^{-28}$ e cm can be reached, {\it Case II} can be tested.
The test of {\it Case I}  is more challenging. But a proposed measurement for proton EDM $d_p$ using proton storage ring can reach a sensitivity of~\cite{proton-edm} $10^{-29}$ e cm which can start to put constraints
on {\it Case I}. While for $d_e$, the current limit already excludes certain
parameter space. Even for $V_{\Sigma 1} V_{h 1}$ as small as $10^{-2}$, {\it Case II}  can reach current bound. For {\it Case I}, to reach current limit of $d_e$, $V_{\Sigma 1} V_{h 1}$ needs
to be close to the maximal which is unlikely. But proposed new experiments will further improve the sensitivity to $O(10^{-30})$ e cm \cite{Kara:2012ay}.
The model can be very well tested.

As discussed in Ref.~\cite{Arguelles:2016ney}, collider studies may probe several other aspects of the model since the new particle masses are $\mathcal{O}$(TeV). A particularly interesting signature involves production of one or two triplet scalars, leading ultimately to a pair of displaced lepton jets in conjunction with one or more prompt objects. The specific final states and corresponding branching ratios can provide information on $\beta_X/\Lambda$, whereas the EDM is sensitive to ${\tilde\beta}_X/\Lambda$. Direct production of the $f$ particles, together with determination of $m_X$ ({\it e.g.}, via measurement of the lepton jet invariant mass) and decay length can provide complementary information. Discovery of  these properties, in combination with a large electron EDM, would provide strong evidence for the triplet $f$ model.
We would also like to point out that the model's new CPV phase $\delta$, together with the extended scalar sector potential that could accommodate a first order electroweak phase transition\cite{Patel:2012pi,Blinov:2015sna,Inoue:2015pza,Niemi:2018asa,Niemi:2020hto}, may provide the ingredients needed to generate the cosmic baryon asymmetry via electroweak baryogenesis.
We will investigate these possibilities in future work.

\section*{Acknowledgments}
This work was supported in part by Key Laboratory for Particle Physics, Astrophysics and Cosmology, Ministry of Education, and Shanghai Key Laboratory for Particle Physics and Cosmology (Grant No. 15DZ2272100), and in part by the NSFC (Grant Nos. 11735010, 11975149, 12090064 and 19Z103010239). XGH was supported in part by the MOST (Grant No. MOST 106-2112-M-002-003-MY3 ). MJRM was also supported in part under
U.S. Department of Energy contract number DE- SC0011095.


\begin{thebibliography}{99}

%\cite{Galison:1983pa}
\bibitem{Galison:1983pa}
P.~Galison and A.~Manohar,
%``TWO Z's OR NOT TWO Z's?,''
Phys. Lett. B \textbf{136} (1984), 279-283.
%doi:10.1016/0370-2693(84)91161-4
%181 citations counted in INSPIRE as of 04 Mar 2021

%\cite{Holdom:1985ag}
\bibitem{Holdom:1985ag}
B.~Holdom,
%``Two U(1)'s and Epsilon Charge Shifts,''
Phys. Lett. B \textbf{166} (1986), 196-198.
%doi:10.1016/0370-2693(86)91377-8
%1802 citations counted in INSPIRE as of 04 Mar 2021

%\cite{Foot:1991kb}
\bibitem{Foot:1991kb}
R.~Foot and X.~G.~He,
%``Comment on Z Z-prime mixing in extended gauge theories,''
Phys. Lett. B \textbf{267} (1991), 509-512.
%doi:10.1016/0370-2693(91)90901-2
%206 citations counted in INSPIRE as of 04 Mar 2021

%\cite{Pospelov:2008zw}
%\bibitem{Pospelov:2008zw}
%M.~Pospelov,
%``Secluded U(1) below the weak scale,''
%Phys. Rev. D \textbf{80} (2009), 095002.
%doi:10.1103/PhysRevD.80.095002
%[arXiv:0811.1030 [hep-ph]].
%688 citations counted in INSPIRE as of 04 Mar 2021

%\cite{Pospelov:2007mp}
%\bibitem{Pospelov:2007mp}
%M.~Pospelov, A.~Ritz and M.~B.~Voloshin,
%``Secluded WIMP Dark Matter,''
%Phys. Lett. B \textbf{662} (2008), 53-61.
%doi:10.1016/j.physletb.2008.02.052
%[arXiv:0711.4866 [hep-ph]].
%811 citations counted in INSPIRE as of 04 Mar 2021

%\cite{Essig:2013lka}
%\bibitem{Essig:2013lka}
%R.~Essig, \textit{et al.}
%J.~A.~Jaros, W.~Wester, P.~Hansson Adrian, S.~Andreas, T.~Averett, O.~Baker, B.~Batell, M.~Battaglieri and J.~Beacham,
%``Working Group Report: New Light Weakly Coupled Particles,''
%[arXiv:1311.0029 [hep-ph]].
%565 citations counted in INSPIRE as of 04 Mar 2021

%\cite{Tulin:2017ara}
%\bibitem{Tulin:2017ara}
%S.~Tulin and H.~B.~Yu,
%``Dark Matter Self-interactions and Small Scale Structure,''
%Phys. Rept. \textbf{730} (2018), 1-57.
%doi:10.1016/j.physrep.2017.11.004
%[arXiv:1705.02358 [hep-ph]].
%410 citations counted in INSPIRE as of 04 Mar 2021

%\cite{Battaglieri:2017aum}
%\bibitem{Battaglieri:2017aum}
%M.~Battaglieri, \textit{et al.}
%A.~Belloni, A.~Chou, P.~Cushman, B.~Echenard, R.~Essig, J.~Estrada, J.~L.~Feng, B.~Flaugher and P.~J.~Fox,
%``US Cosmic Visions: New Ideas in Dark Matter 2017: Community Report,''
%[arXiv:1707.04591 [hep-ph]].
%384 citations counted in INSPIRE as of 04 Mar 2021

%\cite{Chen:2009dm}
\bibitem{Chen:2009dm}
F.~Chen, J.~M.~Cline and A.~R.~Frey,
%``A New twist on excited dark matter: Implications for INTEGRAL, PAMELA/ATIC/PPB-BETS, DAMA,''
Phys. Rev. D \textbf{79} (2009), 063530.
%doi:10.1103/PhysRevD.79.063530
%[arXiv:0901.4327 [hep-ph]].
%48 citations counted in INSPIRE as of 04 Mar 2021

%\cite{Chen:2009ab}
\bibitem{Chen:2009ab}
F.~Chen, J.~M.~Cline and A.~R.~Frey,
%``Nonabelian dark matter: Models and constraints,''
Phys. Rev. D \textbf{80} (2009), 083516.
%doi:10.1103/PhysRevD.80.083516
%[arXiv:0907.4746 [hep-ph]].
%93 citations counted in INSPIRE as of 04 Mar 2021

%\cite{Baek:2013dwa}
%\bibitem{Baek:2013dwa}
%S.~Baek, P.~Ko and W.~I.~Park,
%``Hidden sector monopole, vector dark matter and dark radiation with Higgs portal,''
%JCAP \textbf{10} (2014), 067.
%doi:10.1088/1475-7516/2014/10/067
%[arXiv:1311.1035 [hep-ph]].
%69 citations counted in INSPIRE as of 04 Mar 2021

%\cite{Barello:2015bhq}
\bibitem{Barello:2015bhq}
G.~Barello, S.~Chang and C.~A.~Newby,
%``Correlated signals at the energy and intensity frontiers from non-Abelian kinetic mixing,''
Phys. Rev. D \textbf{94} (2016) no.5, 055018.
%doi:10.1103/PhysRevD.94.055018
%[arXiv:1511.02865 [hep-ph]].
%10 citations counted in INSPIRE as of 04 Mar 2021

%\cite{Arguelles:2016ney}
\bibitem{Arguelles:2016ney}
C.~A.~Arg\"uelles, X.~G.~He, G.~Ovanesyan, T.~Peng and M.~J.~Ramsey-Musolf,
%``Dark Gauge Bosons: LHC Signatures of Non-Abelian Kinetic Mixing,''
Phys. Lett. B \textbf{770} (2017), 101-107.
%doi:10.1016/j.physletb.2017.04.037
%[arXiv:1604.00044 [hep-ph]].
%13 citations counted in INSPIRE as of 04 Mar 2021

%\cite{Fuyuto:2019vfe}
\bibitem{Fuyuto:2019vfe}
K.~Fuyuto, X.~G.~He, G.~Li and M.~Ramsey-Musolf,
%``CP-violating Dark Photon Interaction,''
Phys. Rev. D \textbf{101} (2020) no.7, 075016.
%doi:10.1103/PhysRevD.101.075016
%[arXiv:1902.10340 [hep-ph]].
%6 citations counted in INSPIRE as of 04 Mar 2021

%\cite{Zyla:2020zbs}
\bibitem{Zyla:2020zbs}
P.~A.~Zyla \textit{et al.} [Particle Data Group],
%``Review of Particle Physics,''
PTEP \textbf{2020} (2020) no.8, 083C01.
%doi:10.1093/ptep/ptaa104


\bibitem{type3-seesaw}
R. Foot, H. Lew, X-G He and G. Joshi, Z. Phys. C 44(1989)441.

%\cite{Bhattacharya:2015esa}
%\bibitem{Bhattacharya:2015esa}
%T.~Bhattacharya, V.~Cirigliano, R.~Gupta, H.~W.~Lin and B.~Yoon,
%``Neutron Electric Dipole Moment and Tensor Charges from Lattice QCD,''
%Phys. Rev. Lett. \textbf{115} (2015) no.21, 212002.
%doi:10.1103/PhysRevLett.115.212002
%[arXiv:1506.04196 [hep-lat]].

%\cite{Bhattacharya:2015wna}
%\bibitem{Bhattacharya:2015wna}
%T.~Bhattacharya \textit{et al.} [PNDME],
%``Iso-vector and Iso-scalar Tensor Charges of the Nucleon from Lattice QCD,''
%Phys. Rev. D \textbf{92} (2015) no.9, 094511.
%doi:10.1103/PhysRevD.92.094511
%[arXiv:1506.06411 [hep-lat]].

%\cite{Aad:2020fzq}
\bibitem{Aad:2020fzq}
G.~Aad \textit{et al.} [ATLAS],
%``Search for type-III seesaw heavy leptons in dilepton final states in $pp$ collisions at $\sqrt{s}$ = 13 TeV with the ATLAS detector,''
Eur. Phys. J. C \textbf{81} (2021) no.3, 218.
%doi:10.1140/epjc/s10052-021-08929-9
%[arXiv:2008.07949 [hep-ex]].
%7 citations counted in INSPIRE as of 18 Mar 2021

%\cite{Sirunyan:2019bgz}
\bibitem{Sirunyan:2019bgz}
A.~M.~Sirunyan \textit{et al.} [CMS],
%``Search for physics beyond the standard model in multilepton final states in proton-proton collisions at $\sqrt{s} =$ 13 TeV,''
JHEP \textbf{03} (2020), 051.
%doi:10.1007/JHEP03(2020)051
%[arXiv:1911.04968 [hep-ex]].
%20 citations counted in INSPIRE as of 18 Mar 2021

\bibitem{he-neutron}
X-G He, B. McKellar and S. Pakvasa, Int. J. Mod. Phys. A4 (1989) 5011; A 6 (1991) 1063 (erratum).

%\cite{Engel:2013lsa}
\bibitem{Engel:2013lsa}
J.~Engel, M.~J.~Ramsey-Musolf and U.~van Kolck,
%``Electric Dipole Moments of Nucleons, Nuclei, and Atoms: The Standard Model and Beyond,''
Prog. Part. Nucl. Phys. \textbf{71} (2013), 21-74.
%doi:10.1016/j.ppnp.2013.03.003
%[arXiv:1303.2371 [nucl-th]].

%\cite{Chupp:2017rkp}
\bibitem{michael-edm}
T.E.~Chupp, P.~Fierlinger, M.J.~Ramsey-Musolf and J.T.~Singh,
%``Electric dipole moments of atoms, molecules, nuclei, and particles,''
Rev. Mod. Phys. \textbf{91} (2019) no.1, 015001
%doi:10.1103/RevModPhys.91.015001
%[arXiv:1710.02504 [physics.atom-ph]].


\bibitem{barr-zee}
S. M. Barr and A. Zee, Phys. Rev. Lett. 65(1990)21.


%\cite{Andreev:2018ayy}
\bibitem{Andreev:2018ayy}
V.~Andreev \textit{et al.} [ACME],
%``Improved limit on the electric dipole moment of the electron,''
Nature \textbf{562} (2018) no.7727, 355-360.
%doi:10.1038/s41586-018-0599-8

%\cite{Abel:2020gbr}
\bibitem{Abel:2020gbr}
C.~Abel \textit{et al.} [nEDM],
%``Measurement of the permanent electric dipole moment of the neutron,''
Phys. Rev. Lett. \textbf{124} (2020) no.8, 081803.
%doi:10.1103/PhysRevLett.124.081803
%[arXiv:2001.11966 [hep-ex]].


%\cite{Heister:2002ik}
%\bibitem{Heister:2002ik}
%A.~Heister \textit{et al.} [ALEPH],
%``Search for anomalous weak dipole moments of the tau lepton,''
%Eur. Phys. J. C \textbf{30} (2003), 291-304.
%doi:10.1140/epjc/s2003-01286-1
%[arXiv:hep-ex/0209066 [hep-ex]].

%\cite{Arroyo-Urena:2017sfb}
%\bibitem{Arroyo-Urena:2017sfb}
%M.~A.~Arroyo-Ure\~na, G.~Tavares-Velasco and G.~Hern\'andez-Tom\'e,
%``Weak dipole moments of the tau lepton in models with an extended scalar sector,''
%Phys. Rev. D \textbf{97} (2018) no.1, 013006.
%doi:10.1103/PhysRevD.97.013006
%[arXiv:1711.04327 [hep-ph]].

%\cite{Aad:2020plj}
%\bibitem{Aad:2020plj}
%G.~Aad \textit{et al.} [ATLAS],
%``A search for the $Z\gamma$ decay mode of the Higgs boson in $pp$ collisions at $\sqrt{s}$ = 13 TeV with the ATLAS detector,''
%Phys. Lett. B \textbf{809} (2020), 135754.
%doi:10.1016/j.physletb.2020.135754
%[arXiv:2005.05382 [hep-ex]].

%\cite{Bhattacharya:2015wna}
\bibitem{Bhattacharya:2015wna}
T.~Bhattacharya, V. Cirigliano, S.D. Cohen,R. Gupta, A. Joseph, H.W. Lin, B. Yoon,
%``Iso-vector and Iso-scalar Tensor Charges of the Nucleon from Lattice QCD,''
Phys. Rev. D \textbf{92} (2015) no.9, 094511.
%doi:10.1103/PhysRevD.92.094511
%[arXiv:1506.06411 [hep-lat]].
%109 citations counted in INSPIRE as of 30 Mar 2021

%\cite{Chiang:2020rcv}
\bibitem{Chiang:2020rcv}
C.~W.~Chiang, G.~Cottin, Y.~Du, K.~Fuyuto and M.~J.~Ramsey-Musolf,
%``Collider Probes of Real Triplet Scalar Dark Matter,''
JHEP \textbf{01} (2021), 198
%doi:10.1007/JHEP01(2021)198
%[arXiv:2003.07867 [hep-ph]].
%15 citations counted in INSPIRE as of 04 Apr 2021


%\cite{Leung:2019wao}
\bibitem{Leung:2019wao}
K.~K.~H.~Leung, \textit{et al.},
%M.~Ahmed, R.~Alarcon, A.~Aleksandrova, S.~Bae\ss{}ler, L.~Barr\'on-Palos, D.~H.~Beck, J.~Bessuille, M.~A.~Blatnik and M.~Broering,
%``The neutron electric dipole moment experiment at the Spallation Neutron Source,''
EPJ Web Conf. \textbf{219} (2019), 02005.
%doi:10.1051/epjconf/201921902005
%[arXiv:1903.02700 [nucl-ex]].

%\cite{Ayres:2021hoq}
\bibitem{Ayres:2021hoq}
N.~J.~Ayres \textit{et al.} [n2EDM],
%``The design of the n2EDM experiment: nEDM Collaboration,''
Eur. Phys. J. C \textbf{81} (2021) no.6, 512.
%doi:10.1140/epjc/s10052-021-09298-z
%[arXiv:2101.08730 [physics.ins-det]].



\bibitem{proton-edm}
V. Anastassopoulos et al., Rev. Sci. Instrum. 87 (2016) 115116.


%\cite{Kara:2012ay}
\bibitem{Kara:2012ay}
D.~M.~Kara, \textit{et al.}
%I.~J.~Smallman, J.~J.~Hudson, B.~E.~Sauer, M.~R.~Tarbutt and E.~A.~Hinds,
%``Measurement of the electron's electric dipole moment using YbF molecules: methods and data analysis,''
New J. Phys. \textbf{14} (2012), 103051.
%doi:10.1088/1367-2630/14/10/103051
%[arXiv:1208.4507 [physics.atom-ph]].


%\bibitem{W. C. Griffith}
%W. C. Griffith, Plenary talk at ``Interplay between Particle and Astroparticle physics
%2014".

%\cite{Fitch:2020xvg}
%\bibitem{Fitch:2020xvg}
%N.~J.~Fitch, J.~Lim, E.~A.~Hinds, B.~E.~Sauer and M.~R.~Tarbutt,
%``Methods for measuring the electron EDM using ultracold YbF molecules,''
%[arXiv:2009.00346 [physics.atom-ph]].
%0 citations counted in INSPIRE as of 05 Apr 2021

%
\bibitem{Patel:2012pi}
H.~H.~Patel and M.~J.~Ramsey-Musolf,
%``Stepping Into Electroweak Symmetry Breaking: Phase Transitions and Higgs Phenomenology,''
Phys. Rev. D \textbf{88}, 035013 (2013).
%doi:10.1103/PhysRevD.88.035013
%[arXiv:1212.5652 [hep-ph]].

%\cite{Blinov:2015sna}
\bibitem{Blinov:2015sna}
N.~Blinov, J.~Kozaczuk, D.~E.~Morrissey and C.~Tamarit,
%``Electroweak Baryogenesis from Exotic Electroweak Symmetry Breaking,''
Phys. Rev. D \textbf{92}, no.3, 035012 (2015).
%doi:10.1103/PhysRevD.92.035012
%[arXiv:1504.05195 [hep-ph]].

%\cite{Inoue:2015pza}
\bibitem{Inoue:2015pza}
S.~Inoue, G.~Ovanesyan and M.~J.~Ramsey-Musolf,
%``Two-Step Electroweak Baryogenesis,''
Phys. Rev. D \textbf{93}, 015013 (2016).
%doi:10.1103/PhysRevD.93.015013
%[arXiv:1508.05404 [hep-ph]].
%54 citations counted in INSPIRE as of 11 Nov 2021

%\cite{Niemi:2018asa}
\bibitem{Niemi:2018asa}
L.~Niemi, H.~H.~Patel, M.~J.~Ramsey-Musolf, T.~V.~I.~Tenkanen and D.~J.~Weir,
%``Electroweak phase transition in the real triplet extension of the SM: Dimensional reduction,''
Phys. Rev. D \textbf{100}, no.3, 035002 (2019).
%doi:10.1103/PhysRevD.100.035002
%[arXiv:1802.10500 [hep-ph]].

%\cite{Niemi:2020hto}
\bibitem{Niemi:2020hto}
L.~Niemi, M.~J.~Ramsey-Musolf, T.~V.~I.~Tenkanen and D.~J.~Weir,
%``Thermodynamics of a Two-Step Electroweak Phase Transition,''
Phys. Rev. Lett. \textbf{126}, no.17, 171802 (2021).
%doi:10.1103/PhysRevLett.126.171802
%[arXiv:2005.11332 [hep-ph]].

\end{thebibliography}
\end{document}